\documentclass[aps,prb,twocolumn,reprint,amsmath,amssymb]{revtex4-1}

\usepackage{graphicx, bm}
\usepackage{fp}
\usepackage{siunitx}
\usepackage{url}
\usepackage{hyperref}
\newcommand{\fig}[1]{Fig.~\ref{fig:#1}}
\newcommand{\eqn}[1]{Eqn.~\ref{eqn:#1}}
\newcommand{\geff}{\ensuremath{g_\mathrm{eff}}}
\newcommand{\omegac}{\ensuremath{\omega_\mathrm{c}}}

\newcommand{\omegas}{\ensuremath{\omega_\mathrm{s}}}

\newcommand{\gud}{\ensuremath{g^{\uparrow\downarrow}}}
\usepackage[sort&compress]{natbib}

\begin{document}

\title{Spin pumping in strongly coupled magnon-photon systems}

\author{H. Maier-Flaig$^{1,2 }$\footnote{Electronic address: hannes.maier-flaig@wmi.badw.de}, M. Harder$^{3}$, R. Gross$^{1,2,4}$,  H. Huebl$^{1,2,4}$, S. T. B. Goennenwein$^{1,2,4}$}

\affiliation{$^{1}$ Walther-Mei\ss ner-Institut, Bayerische Akademie der Wissenschaften, 85748 Garching, Germany}
\affiliation{$^{2}$ Physik-Department, Technische Universit\"{a}t M\"{u}nchen, 85748 Garching, Germany}
\affiliation{$^{3}$ Department of Physics and Astronomy, University of Manitoba, Winnipeg, Canada R3T 2N2}
\affiliation{$^{4}$ Nanosystems Initiative Munich, Schellingstra\ss e 4, D-80799 M\"{u}nchen, Germany}

\date{\today}

\begin{abstract}

We experimentally investigate magnon-polaritons, arising in ferrimagnetic resonance experiments in a microwave cavity with a tuneable quality factor. To his end, we simultaneously measure the electrically detected spin pumping signal and microwave reflection (the ferrimagnetic resonance signal) of a yttrium iron garnet (YIG) / platinum (Pt) bilayer in the microwave cavity.  The coupling strength of the fundamental magnetic resonance mode and the cavity is determined from the microwave reflection data. All features of the magnetic resonance spectra  predicted by first principle calculations and an input-output formalism agree with our experimental observations. By changing the decay rate of the cavity at constant magnon-photon coupling rate, we experimentally tune in and out of the strong coupling regime and successfully model the corresponding change of the spin pumping signal.  Furthermore, we observe the coupling and spin pumping of several spin wave modes and provide a quantitative analysis of their coupling rates to the cavity.

\end{abstract}


\maketitle

\section{Introduction}
Motivated by the vision of hybrid quantum information systems combining the fast manipulation rates of superconducting qubits and the long coherence times of spin ensembles, strong spin-photon coupling is a major goal of quantum information memory applications.  Coherent information exchange between microwave cavity photons and a spin ensemble was initially demonstrated for paramagnetic systems \cite{Schuster2010,Kubo2010,Zollitsch2015}, but only recently has this concept been transferred to magnetically ordered systems, where coupling rates of hundreds of megahertz can be achieved.\cite{Soykal2010,Huebl2013,Tabuchi2014,Zhang2014}  Utilizing the flexibility of exchange coupled magnetically ordered systems, more complex architectures involving multiple magnetic elements have already been developed\cite{Zhang2015,Lambert2015}. Additionally, magnetically ordered systems allow to study classical strong coupling physics even at room temperatures.\cite{Huebl2013,Tabuchi2014,Zhang2014,Zhang2015,Lambert2015,Yao2015}

Moreover, a key advantage of magnetically ordered systems over their paramagnetic counterparts -- which has yet to be fully explored -- is the ability to probe magnetic excitations electrically through spin pumping and the inverse spin Hall effect. Spin pumping, in general, relies on ferromagnet-normal metal (FM/NM) heterostructures and has been demonstrated for a wide variety of material combinations\cite{Czeschka2011a}. Under resonant absorption of microwaves, the precessing magnetisation in the ferromagnet sources a spin current into the normal metal, where it is converted into a charge current via the inverse spin Hall effect. This spin Hall charge current is then  detected. In ferromagnetic insulator (FMI)-based FMI/NM heterostructures, charge current signals from the rectification of the microwave electric field are very small\cite{Iguchi2014}, leading to a dominant spin pumping/spin Hall signal.  This has led to much research on FMI/NM heterostructures, of which the Yttrium Iron Garnet (YIG)/Platinum(Pt) bilayers we use are a prime example. Spin pumping is a well understood effect for weak photon-magnon coupling \cite{Tserkovnyak2002,Czeschka2011a}, i.e. for situations where the decay rates of the cavity and the magnetic system are larger than the photon-magnon coupling strength. However, the large spin density of YIG and the resulting large effective coupling strength allows one to reach the strong coupling regime also in typical spin pumping experiments. The experimental observation\cite{Bai2015} and theoretical treatment   \cite{Cao2015,Lotze2015} of spin pumping in a strongly coupled magnon-photon system has only recently been performed. These results suggest that combining spin pumping and strong magnon-photon coupling may enable the transmission and electrical read out of quantum states in ferromagnets using a hybrid architecture. Experiments directly linking spin pumping in the weak and strong coupling regime are, however, still missing. Such experiments are one important step towards understanding the functional principle and key requirements for such a hybrid architecture.

In this paper, we present a systematic study of the magnon-photon coupling in magnetic resonance experiments in a YIG/Pt bilayer mounted in a commercially available EPR cavity. We measure both the microwave reflection spectra and the electrically detected spin pumping signal in the system.  The tuneable cavity quality allows us to systematically move in and out of the strong coupling regime. Measurements with high magnetic field and frequency resolution allow us to clearly observe the coupling of spin wave modes with the hybridized cavity--fundamental FMR mode. We explore a different approach as recently used by \citet{Zhang2014}: In our setup, instead of tuning the cavity frequency we tune its decay rate while the effective magnon-photon coupling rate and the magnon decay rate stay constant. We thus achieve a transition from the strongly coupled regime where the decay rates of spin and cavity system are both considerably smaller than the effective coupling rate, to the weakly coupled regime where the cavity decay rate is much higher than the magnon-photon coupling rate. This regime is also called the regime of magnetically induced transparency (MIT)\cite{Zhang2014}.

This paper is organized as follows: In Sec. \ref{sec:theory} we review the general theory of the coupled magnon-photon system and the main features of spin pumping in the case of strong coupling.  In Sec. \ref{sec:exp} we describe the experimental details of recording the microwave reflection of the system as a function of frequency and applied magnetic field while simultaneously recording the DC spin pumping voltage across the Pt.  Finally in Sec. \ref{sec:discuss} we present our observation of strong coupling between the cavity mode and both the fundamental magnetic resonance and standing spin wave modes.  We also demonstrate the transition from strong to weak coupling by tuning the cavity line width and discuss the difference in the experimental spin pumping signature in both the strong and weak regimes.

\section{Theory}
\label{sec:theory}
\subsection{Photon-Magnon Dispersion}
Conventionally, ferromagnetic resonance (FMR) is modeled in terms of the Landau-Lifshitz-Gilbert (LLG) equation which describes the dynamics of a magnetic moment in the presence of a magnetic field.  In a static magnetic field $H_0$, the magnetic moment will precess with the Larmor frequency $\omega_\mathrm{s}$. In detail, $\omega_\mathrm{s}$ depends on the static field strength and on its orientation due to anisotropy\cite{Brandlmaier2008}. This precessional motion can be resonantly excited by a time varying microwave magnetic field $H_1$ with a frequency close to $\omega_\mathrm{s}$. To observe spin pumping in FM/NM heterostuctures, the field $H_0$ should be applied perpendicular to the surface normal (i.e. in the interface plane)\cite{Heinrich2003, Tserkovnyak2002,Mosendz2010,Czeschka2011a}. In this case, the FMR dispersion (in the absence of crystalline magnetic anisotropy) is $\omega_s =  \gamma \mu_0 \sqrt{ H_0 \left( H_0 + M_\mathrm{s}\right) }$\cite{Kittel2005}. Here, $M_\mathrm{s}$ is the material specific saturation magnetization, $\gamma$ is the material specific gyromagnetic ratio and $\mu_0$ is the vacuum permeability. In the limit $H_0 \gg M_\mathrm{s}$   the resonance frequency is thus linear in magnetic field. Contrary to the spin resonance frequency $\omegas$, the resonance frequency $\omegac$ of a macroscopic cavity is determined by geometrical and dielectric parameters only and therefore does not depend on the magnetic field.  However, since the magnonic and the photonic mode interact in resonance, we expect modifications to the pure FMR and pure cavity dispersions. To be specific, we will observe an anticrossing of the FMR and the cavity dispersion for a sufficiently strong magnon-photon coupling.

To describe the coupling between the cavity mode and the spin excitation the quantum mechanical Tavis-Cummings model \cite{Tavis1968,Fink2009} and classical first principles\cite{Cao2015} approaches using the input-output formalism\cite{Huebl2013} have successfully been used. For the dipolar interaction assumed in the models, the single spin-single photon coupling strength $g_0$ is proportional to the vacuum microwave magnetic field $H_1^0$ and the dipole moment $m$ of the spin. In the scope of the Tavis-Cummings model, it has been shown that the collective coupling strength $\geff$ to an ensemble of spins is proportional to the square root of the number of polarized spins for the coupling to the vaccum microwave magnetic field. In a classical theory, \citet{Cao2015} derived that this $\sqrt{N}$ behaviour prevails also for the magnon-photon coupling in magnetically ordered systems. Here, the total magnetization and thus the filling factor of the ferromagnetic material in the cavity, can be used as a measure for the total number of spins. 

The characteristic fingerprint of strong coupling is the formation of an observable anti-crossing of the cavity and the spin dispersion relation close to resonance. Note, that the presence of stong coupling and the accompanied visible anti-crossing of the dispersion relations requires that the effective coupling \geff\ exceeds the loss rates of the spins ($\gamma_\mathrm{s}$) and the cavity ($\kappa_\mathrm{i}+\kappa_\mathrm{e}$). Experimentally, we tune the spin resonance frequency $\omegas$ across the cavity resonance frequency $\omegac$ via an externally applied static magnetic field. The coupled system can most simply be modelled in the vicinity of the resonance frequency using two coupled harmonic oscillators, where the resonance frequency is\cite{Huebl2013}:
\begin{equation}
  \label{eqn:HO}
  \omega_\pm = \omega_\mathrm{c} + \frac \Delta 2 \pm \frac 1 2 \sqrt{\Delta^2 + 4 g_\mathrm{eff}^2}
\end{equation}
Here, $\Delta = \gamma \left( \mu_0 H_0 - \mu_0 H_\mathrm{res} \right)$ is the spin-cavity detuning with $H_\mathrm{res}$ statisfying the spin resonance condition for a given cavtiy frequency $\omega_\mathrm{c}$. 

In ferromagnetic films, additional magnetic modes, so-called perpendicular standing spin waves modes, appear due to magnetic boundary conditions. For the condition where the magnetization is pinned at least at one surface of the film (and in the absence of any anisotropies or magnetic gradients) the magnon spectrum can easily be calculated\cite{Kittel2005}. The difference of the resonance field of the $n$\textsuperscript{th}-mode from the fundamental mode $H^n_\mathrm{res}-H^1_\mathrm{res}$ is proportional to $n^2$. \citet{Cao2015} also calculated the expected coupling strength for different modes and found that the coupling decreases with increasing mode number as $\geff \propto 1/n$. This can be understood when considering the microwave mode profiles and the fact that the spatial mode profile of the microwave field $H_1^0$ in a cavity is typically homogeneous and in phase throughout the thickness of the (thin film) sample. Therefore only every second mode can be excited and the effective magnetization to which the microwave can couple to is reduced to $\frac M n$.
\subsection{Spin pumping and strong coupling}
Spin pumping in ferromagnet/normal metal bilayers in the weak coupling regime is well understood\cite{Tserkovnyak2002,Czeschka2011a,Mosendz2010}: An additional mechanism which damps the magnetization precession becomes available by spin pumping, as the precessing magnetisation is driving a spin current into the adjacent normal metal.\cite{Tserkovnyak2002} In electrically detected spin pumping, this spin current is then converted into a charge current via the inverse spin Hall effect (ISHE). For electrical open circuit conditions, one thus obtains a voltage which scales as\cite{Mosendz2010,Czeschka2011a} $V_\mathrm{SP} \propto g^{\uparrow\downarrow}\lambda_\mathrm{SD}\tanh{\frac{t_\mathrm{N}}{2\lambda_\mathrm{SD}}}\sin^2{\theta}$. It, thus, contains information on the spin mixing conductance $g^{\uparrow\downarrow}$, the spin diffusion length $\lambda_\mathrm{SD}$, the magnetization precession cone angle $\theta$ and depends on the thickness of the normal metal layer $t_\mathrm{N}$. The maximal precession cone angle $\theta$ and thus the maximal expected spin pumping voltage depends on the microwave power but also on the coupling strength between cavity and spin system. For strong coupling, the cone angle is expected to be reduced as compared to the weak coupling case due to the hybridized nature of the excitation at its maximal intensity.

The other contributions in the equation for $V_\mathrm{SP}$ are material constants: The spin mixing conductance $\gud$ describes the the transparency of the ferromagnet/normal metal interface und limits the spin pumping efficiency generally; the spin diffusion length $\lambda_\mathrm{SD}$ in conjunction with the normal metal thickness $t_\mathrm{N}$ accounts for a spin  accumulation in the normal metal and reduces the spin pumping efficiency if $t_\mathrm{N} \lessapprox \lambda_\mathrm{SD}$.

\section{Experimental details}
\label{sec:exp}
\begin{figure}
\includegraphics[width=0.4\textwidth]{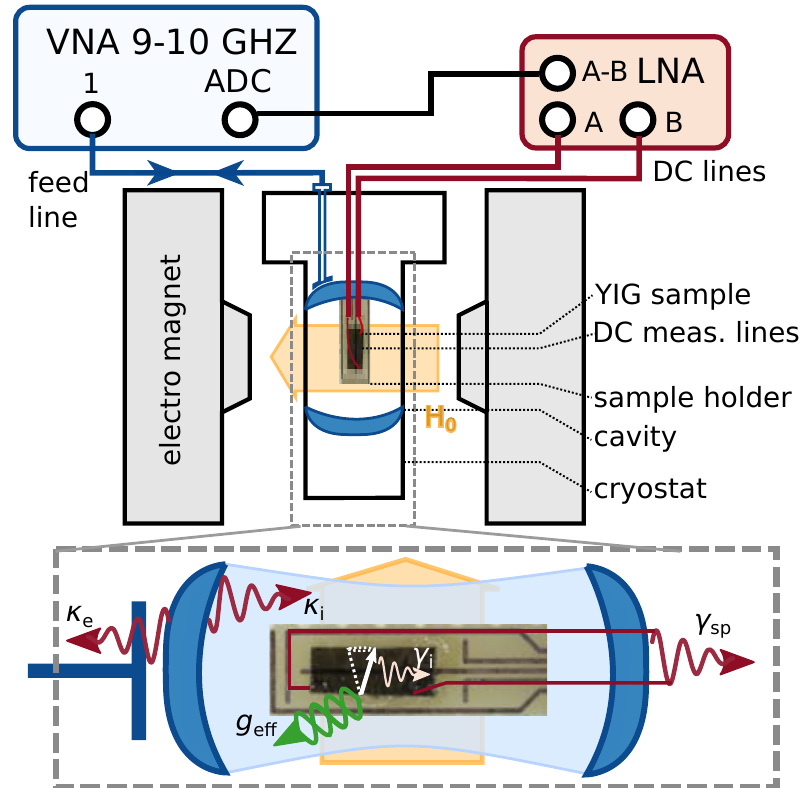}
\caption{ Block diagram of the experimental setup and sample mounting. \textbf{(Inset)} Schematic of the coupling scheme illustrating cavity decay due to intrinsic losses viz. losses to the feedline $\kappa_\mathrm{c} = \kappa_\mathrm{i} + \kappa_\mathrm{e}$, spin system decay consisting of intrinsic damping and spin pumping damping $\gamma_\mathrm{i}= \gamma_\mathrm{s} + \gamma_\mathrm{sp}$ as well the collective coupling rate $g_\mathrm{eff}$}
\label{fig:setup}
\end{figure}

\subsection{Sample preparation}
In our experiments we used YIG/Pt heterostructures grown by liquid phase epitaxy on (111)-oriented Gadolinum Gallium Garnet (GGG) substrates. The YIG film thickness was \SI{2.8}{\micro\meter}. In order to produce a high quality interface between YIG and Pt, and thus a large spin mixing conductance  $g^{\uparrow\downarrow}$, we followed the work of \citet{Jungfleisch2013} and first treated the YIG surface by piranha etching for 5~minutes in ambient conditions. Thereafter, the sample was annealed at \SI{500}{\celsius} for 40~minutes in an oxygen atmosphere of \SI{25}{\micro\bar}. Under high vacuum, it was then transferred into an Electron Beam Evaporation (EVAP) chamber where \SI{5}{\nano\meter} Pt was deposited. The exact Pt thickness was determined using X-Ray reflectometry.  However, we note that for our analysis the Pt layer thickness is of minor importance as it was consistently larger than the spin diffusion length $\lambda_\textrm{SD}$ of Pt such that the Pt layer simply serves as a perfect spin sink.

In order to achieve collective strong coupling between magnons and cavity photons, the number of magnetic moments must be sufficiently high. Therefore, we diced the sample into several pieces of different lateral dimensions. Magnetic resonance experiments in the strong coupling regime showed that the $\sqrt{N}$ scaling of the coupling strength discussed in Section~\ref{sec:theory} is indeed obeyed upon comparing samples with different volume and thus different total magnetic moment. In the following, we will focus on a sample with lateral dimensions \SI{2x3}{\milli\meter} which, with the effective spin density $\rho_\textrm{S} = 2.1 \times 10^{22}\,\frac{\mu_\mathrm{B}}{\mathrm{cm}^3}$ of iron atoms in YIG\cite{Gilleo1958}, contains on the order of $4\times10^{17}$ spins. 
Finally, the sample was mounted on a PCB sample carrier and wire bonded as depicted in the inset of \fig{setup}. The carrier itself was mounted on a sample rod which allowed the sample to be accurately positioned in the electrical field node of a Bruker Flexline MD5 dielectric ring cavity in an Oxford Instruments CF935 gas flow cryostat. Shielded DC cabling allowed for the measurement of the ISHE voltage. The detailed design blueprints of the sample rod and chip carrier can be retrieved online\footnote{The designs of the sample holder are published under the CERN Open Hardware License \texttt{http://hannes.maier-flaig.de/flexline-sample-rod}}.

\subsection{Experimental setup}

The Bruker cavity exhibits a TE$_{011}$ mode with an electric field node at the sample position. Its quality factor $Q = \omega / \Delta \omega_\mathrm{c}^\mathrm{FWHM}$ ($\Delta \omega_\mathrm{c}^\mathrm{FWHM}$ being the full width half maximum line width of the cavity) is dominated by the dissipative losses in the dielectric and its finite electrical resistance ($\kappa_\mathrm{i}$) as well as radiation back into the cavity feed line ($\kappa_\mathrm{c}$). By changing the cavity's coupling ratio to the feed line, unloaded coupled quality factors $Q_\mathrm{c}$ from 0 to 8000 can be achieved. This allows tuning in and out of the strong coupling regime easily. Using the gas flow cryostat, different temperatures can be stabilized. All the following experiments have, however, been performed at room temperature.

To measure ferromagnetic resonance (FMR) the cavity was connected to the port of an Agilent N5242A vector network analyzer (VNA). The driving power of $15$~dBm excites at maximum on the order of $N_\mathrm{Ph} = 1.3\times10^{14}$ photons in the cavity which is considerably smaller than the number of spins in the sample ($4\times10^{17}$). In this case, the  theory presented in Sec.~\ref{sec:theory}\  is well justified\cite{Chiorescu2010}. The frequency dependent cavity reflection $S_{11}$ was measured while sweeping the external field $\mu_0 H$ that is created by a water cooled electromagnet. The IF bandwidth was chosen to be \SI{100}{\hertz} which leads to a frequency sweep time of approximately \SI 2 \second\ for each magnetic field step. A calibration of the microwave leads up to the resonator's SMA connector was performed. The calibration did not include the feed line inside the resonator mount, which gave rise to a background signal in the reflection parameter. However, by utilizing the full complex S-parameter for the background subtraction with the inverse mapping technique outlined by \citet{Petersan1998} and a subsequent Lorentzian fit to the magnitude, a reliable measurement of $Q$ is still possible, even for a completely uncalibrated setup. We note that even though standing waves in the mirowave feed line will not appear in the calibrated reflection measurement they will still change the total power in the cavity and therefore may complicate the electrically detected DC spin pumping signal. Uncalibrated measurements did not show sharp feed line resonances in the frequency range studied here but only smooth oscillations with an amplitude of less than \SI{1}{\decibel} and there was no correlation in the DC signal resolved. In order to fit the data and as it improves clarity, we only discuss calibrated measurements in the following.

The DC voltage from the sample was measured along the cavity axis and thus perpendicular to the external magnetic field and the sample normal. It was amplified with a differential voltage amplifier model 560 from Stanford Research Systems. The amplifier was operated in its low noise (4~nV/$\sqrt{\textrm{Hz}}$) mode and set to a gain of $2\times 10^4$. The analog high-pass filter of the amplifier was disabled, however, a low-pass filter with a 6dB roll-off at \SI{1}{kHz} was employed. Limiting the bandwidth of the amplifier by filtering is required in order to achieve a good signal-to-noise ratio. Care has, however, to be taken as the lineshape may be quickly distorted by inappropriate settings and thus the signature of spin pumping might be masked. High-pass filtering can easily lead to a dispersive like contribution to the signal, whereas low-pass filtering will give rise to asymmetric line shapes depending on the ratio of IF bandwidth and low-pass frequency. We made sure that no such distortions contribute to the presented measurements. The amplified voltage signal was finally recorded using the auxiliary input of the VNA simultaneously with the cavity reflection $S_{11}$.

\section{Results and discussion}
\label{sec:discuss}
\begin{figure}
\includegraphics[width=0.45\textwidth]{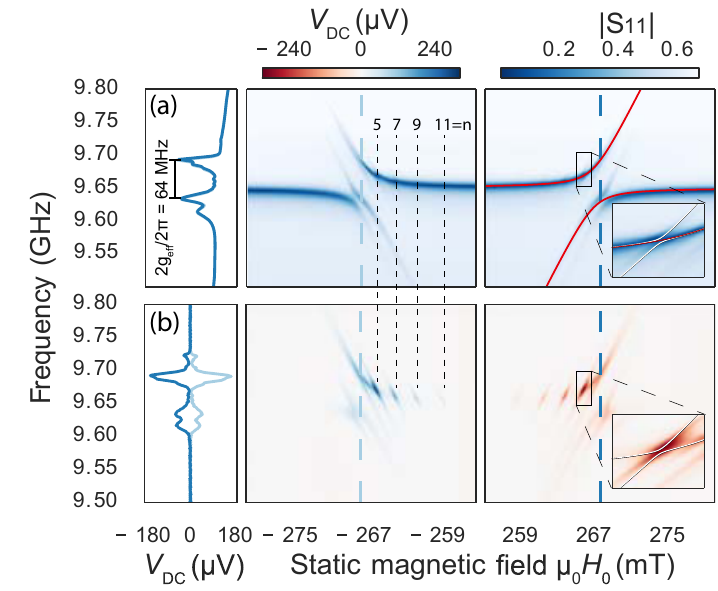}
\caption{\textbf{(a)} Reflection parameter $S_{11}$ recorded while sweeping the magnetic field. Strong coupling of the collective spin excitations is indicated by a clear anticrossing, spin wave modes to the low field side of the main resonance are visible. Black numbers indicated the spin wave mode number. \textbf{(b)} Simultaneously recorded DC voltage. Fundamental and spin wave modes are visible where  the latter couple less strongly and thus pump spin current more efficiently. Insets: Detail of n=5 spin wave mode including the dispersion relation of the strong coupling between the fundamental FMR mode and the cavity as solid red line and the anti-crossing of this hybrid and the spin wave mode (\#5) as white lines. }
\label{fig:posneg}
\end{figure}

We first focus on the case of the so-called critical coupling of the feed line to the cavity in which most FMR experiments are conducted. In this case, the internal loss rate of the cavity equals the loss rate to the feed line and the quality factor is $Q_\mathrm{c} = Q_\textrm{internal}/2$.
Note that inserting a sample and holder into the cavity will reduce the cavity $Q$ by an amount which depends on the sample and holder details such as conductivity and dielectric losses. Based on our measured loaded $Q_\mathrm{c}=706$, the cavity decay rate is calculated to be $\kappa_\mathrm c / 2 \pi = \frac{\omega_\mathrm{r}}{2\pi} /2Q_\mathrm{c} = \SI{6.8}{\mega\hertz}$.

Strong coupling of the magnon and cavity system manifests itself in a characteristic anti-crossing of the (magnetic field independent) cavity resonance frequency and the magnon dispersion that is (approximately) linear in magnetic field. This anti-crossing corresponding to two distinct peaks in a line cut at the resonance field, are immediately visible in the reflection spectrum in \fig{posneg}. The minimal splitting gives the collective coupling strength $g_\textrm{eff}/2\pi = \SI{31.8}{\mega\hertz}$ of the fundamental mode to the cavity. Taking into account the number of spins in the sample, the single spin coupling rate is on the order of $g_0/2\pi=\SI{0.1}{Hz}$ which is in agreement with experiments on paramagnetic systems\cite{Abe2011}.

In our setup, even the coupling of higher order spin wave modes to the cavity can be resolved. We number the spin waves as noted in \fig{posneg} taking into account that with an uniform driving  field only odd modes can be excited. Analysis of the resonance position of the spin wave modes reveals that $H^n_\mathrm{res}-H^1_\mathrm{res}$ in our sample is proportional to $n$ rather than $n^2$. This indicates a non-square like potential well.  Similarly, complicated mode splittings have been reported in literature\cite{Goennenwein2003}. The lowest order spin wave mode that can be easily observed in our setup is shown in the inset of \fig{posneg}~(a) in detail. It exhibits the largest effective coupling (\SI{3}{\mega\hertz}) of all spin wave modes. The red and white lines in Fig.~\ref{fig:posneg}~(a) correspond to the harmonic-oscillator model (\eqn{HO}) for the fundamental mode and the lowest order spin wave mode, respectively. As the spin wave couples to an already hybridized sytem, we superimposed the dispersion $\omega_c = \omega_r \left( B \right)$ of the hybridized system of fundamental mode and unperturbed cavity as the "cavity" mode in the modelling of the spin wave mode couplings.

In order to quantify the coupling strength of the higher order modes which only interact weakly with the hybridized cavity--fundamental FMR mode, we follow the approach of \citet{Herskind2009}. For each field, we fit a Lorentzian to the magnitude of the cavity absorption. From this fit we extract the resonance frequency $\omega_c$ and the half width half maximum of the absorption  $\Delta \omega$ which, in
the weakly coupled spin waves reads as\cite{Herskind2009}
\[
\Delta \omega = \Delta \omega_c + g_\mathrm{eff} \gamma_s / \left( \gamma_s ^2 + \Delta ^2 \right).
\]

The coupling of the spin waves to the already hybridized cavity resonance decreases with the order of the mode. This can be understood by taking into account that the effective magnetization to which the homogeneous microwave field can couple decreases with increasing mode number. The extracted values, $g_{n=7,9,11,13} = \left[3.65, 2.49, 1.64, 1.16\right] \SI{}{\mega\hertz}$, match accurately with the expected $\frac 1 n$ dependence of the coupling strength\cite{Cao2015}.

We attribute the pronounced feature that is seen to the right of the anti-crossing to an unidentified spin wave mode. A similar feature was found in other experiments\citep{Bai2015}  and has been interpreted in the same manner. In our data, we can clearly distinguish between the fundamental mode and this additional mode -- simply by remembering that the relative intensity and coupling strength is expected to be higher for the fundamental mode. Possible origins for the additional mode are an inhomogeneous sample or a gradient in the magnetic properties across the film thickness\cite{Hoekstra1977}. This would be consistent with the unusual spin wave mode splitting. Lastly, we note that the recorded signal in the reflection parameter is completely symmetric upon magnetic field reversal.

The simultaneously recorded DC voltage is shown in \fig{posneg}~(b). Contrary to the reflection parameter, the voltage signal reverses sign on reversing $\mu_0 H_0$. The lineshape that we record for all modes is completely symmetric as far as they can be clearly distinguished from each other. We thus conclude that we observe a signal purely caused by spin pumping and not by any rectifying effect. In a FMI/NM bilayer ($\rho_\mathrm{YIG} \geq \SI{10}{\giga\ohm\meter}$) \cite{Tucciarone1984} rectification can only arize from a change of the spin Hall magnetoresistance (SMR) in the normal metal. According to model calculations \cite{Iguchi2014} this effect is negligible for the system we investigate because of the small magnitude of the SMR effect ($< 0.1 $\%). This notion is further corroborated by the fact that the change in lineshape expected for rectification type signals is not visible in our data. Apart from the spin wave modes which are clearly resolved in the DC voltage signal, we can also clearly see the electrically detected spin pumping voltage originating from the hybridized system of cavity and fundamental FMR mode (the main anti-crossing). The hybridized cavity eigenmodes can, however, pump spin current into the normal metal only very inefficiently and thus the DC voltage we observe is very low.

\begin{figure*}
\includegraphics{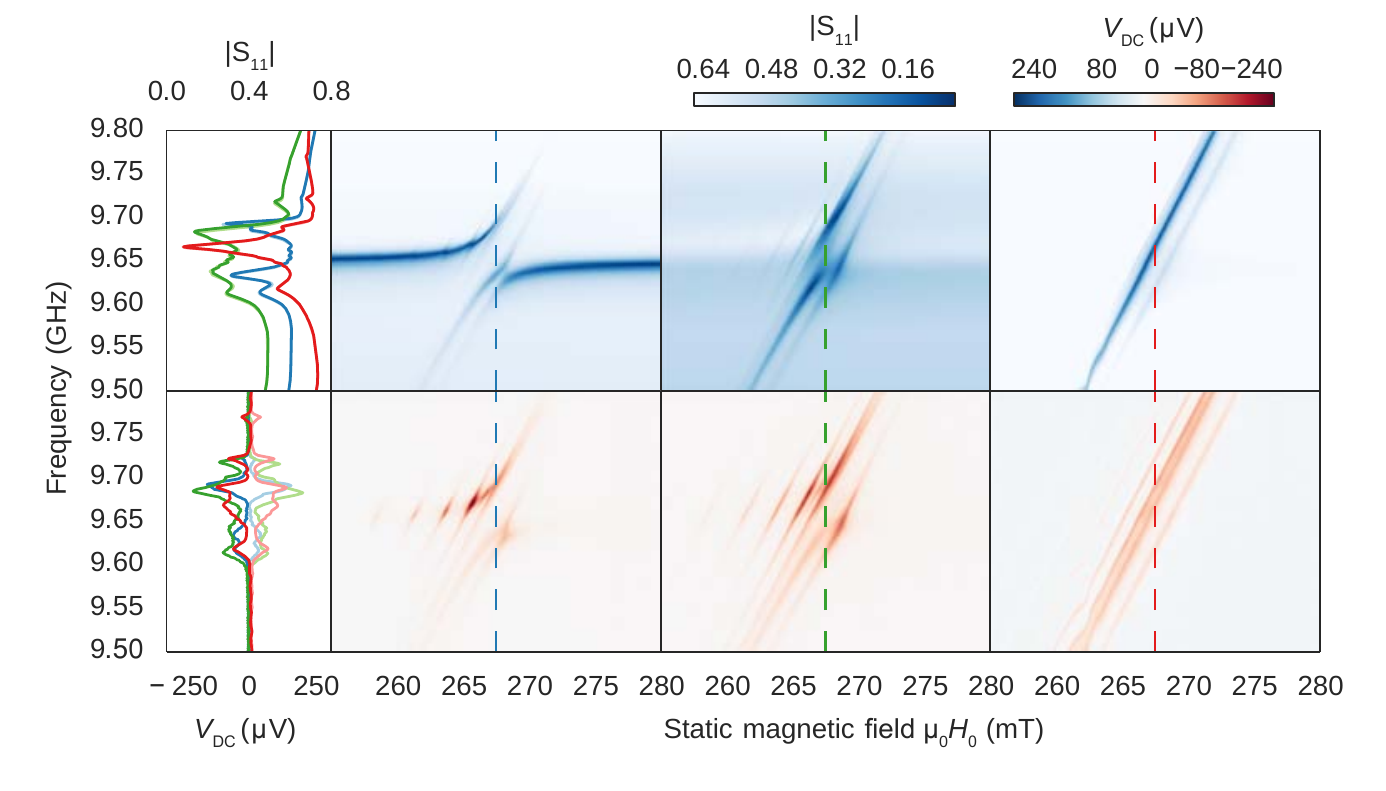}
\caption{Increasing the coupling of the cavity to the feed line (from left to right) increases the cavity loss rate $\kappa_\mathrm{c}$ and thus line width $\Delta \omega/2\pi$. This enables experimental control of the transition between strong and weak coupling.  The line cuts at positive field (intense colors) and negative field (pale colors) again confirm the symmetry, $V(-H_0) = -V(H_0)$ and $S_{11}(-H_0) = S_{11}(H_0)$ and also show the merging of the two dispersion curves during the strong/weak transition.}
\label{fig:coupling}
\end{figure*}

The upper panels of \fig{coupling} show the change in cavity reflection as we gradually increase the coupling of the cavity to the feed line and thus increase the cavity decay rate.  Starting from the critically coupled case (internal cavity losses are equal to losses into the feedline) in the left panel to a highly overcoupled cavity (losses into the cavity feed line dominate the cavity's decay rate) in the right panel, we clearly see an increase in the cavity linewidth up to the point were the unperturbed cavity is no longer recognizable. Correspondingly, the cavity decay rate increases from left to right and, in turn, the microwave magnetic field strength $H_1$ in the cavity decreases. 
For the already weakly coupled spin wave modes the spin pumping voltage decreases with decreasing microwave magnetic field strength $H_1$ resp. available microwave power(indicated by the higher $S_{11}$ parameter) in the cavity.
The DC spin pumping voltage amplitude corresponding to the fundamental mode (lower panels of \fig{coupling}) does, however, not decrease for lower $Q$-factors but stays approximately constant. Considering that the absorbed power of the cavity-spin system stays approximately constant when changing the cavity decay rate as can easily be seen in the line cuts in the upper panels of \fig{coupling} this behaviour can also be understood.

\begin{figure}
\includegraphics{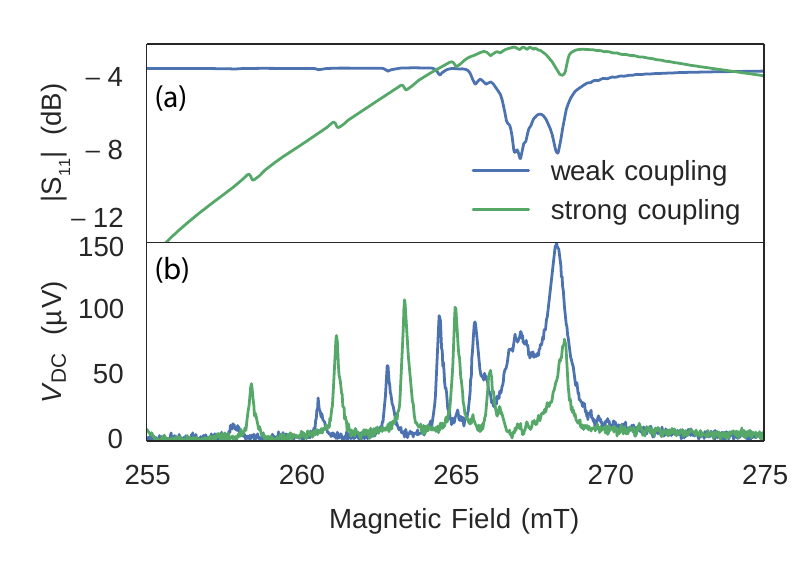}
\caption{Line cuts of (a) reflection parameter and (b) DC voltage at the resonator frequency $\omegac\left( H_0 = 0 \right)$. In the strongly coupled magnon-photon case (green lines), the fundamental mode vanishes as opposed to the weakly coupled case where the magnon spectrum is accurately reproduced}
\label{fig:linecuts}
\end{figure}

The best measure of the true magnon spectrum and line widths of the spin system can be extracted from the highly overcoupled case (right panels of \fig{coupling} and \fig{linecuts}). There, the magnon-photon coupling is negligible compared to the cavity loss rate and therefore, the magnon-cavity mode hybridization does not distort the line shape. A mode that strongly couples with the cavity, on the contrary, can vanish completely in the fixed-frequency spectrum. We finally note that we observe the described anti-crossing due to the magnon-photon coupling and thus the distortion of the lines in a fixed-frequency experiment (with the cavity tuned to high Q, as usually done in cavity-based FMR experiments) already for sample volumes as small as $V=2\times 10^{-3}\SI{} {\cubic\milli\meter}$ in the case of YIG ($M_\mathrm{S} = \SI{140}{\kilo\ampere\per\meter}$). These sample volumes are easily achieved for LPE grown samples and we note that in most cavity FMR experiments\cite{Kajiwara2010} the effects of the coupling need to be taken into account in order to yield accurate results especially when automatic frequency control is employed.

\section{Conclusions}
In summary, we presented systematic measurements of spin pumping in different regimes of the magnon-photon coupling strength. For the fundamental mode of a YIG/Pt bilayer, strong coupling with an effective coupling strength of $g_{\rm eff}/2\pi = \SI{31.8}{\mega\hertz}$ has been achieved at room temperature in a standard EPR cavity. The characteristics of the coupled magnon-photon system fit well to the established theory and are consistent with recent results. Simultaneously, we recorded the electrically detected spin pumping signal of the fundamental mode. We were able to tune the system from the strong to the weak coupling regime by changing the cavity's decay rate. The evolution of the spin pumping signal of the fundamental mode has been analyzed qualitatively and follows the predictions of \citet{Lotze2015}: In the strongly coupled magnon-photon system the spin pumping efficiency is reduced as the precession cone angle is smaller than in the weakly coupled case.  Additionally, we were able to observe coupling and electrically detected spin pumping of several spin wave modes with distinctly different coupling strengths and observe for the first time their $1/n$ dependence predicted by \citet{Cao2015}. Furthermore, we directly demonstrated the implications of strong coupling on fixed-frequency FMR experiments. We conclude that small sample volumes or an highly overcoupled cavity are mandatory for a qualitatively and quantitatively correct evaluation of the magnon spectrum.

\section*{\uppercase{ Acknowledgements}}
We thank Christoph Zollitsch and Johannes Lotze for many valuable discussions and Michaela Lammel for assistance in the sample preparation. M. Harder acknowledges support from the NSERC MSFSS program. We gratefully acknowledge funding via the priority programme Spin Caloric Transport (spinCAT) of Deutsche Forschungsgemeinschaft (Project GO 944/4), SFB 631 C3 and the priority programm SPP 1601 (HU 1896/2-1).

\bibliography{references}

\begin{thebibliography}{33}%
\makeatletter
\providecommand \@ifxundefined [1]{%
 \@ifx{#1\undefined}
}%
\providecommand \@ifnum [1]{%
 \ifnum #1\expandafter \@firstoftwo
 \else \expandafter \@secondoftwo
 \fi
}%
\providecommand \@ifx [1]{%
 \ifx #1\expandafter \@firstoftwo
 \else \expandafter \@secondoftwo
 \fi
}%
\providecommand \natexlab [1]{#1}%
\providecommand \enquote  [1]{``#1''}%
\providecommand \bibnamefont  [1]{#1}%
\providecommand \bibfnamefont [1]{#1}%
\providecommand \citenamefont [1]{#1}%
\providecommand \href@noop [0]{\@secondoftwo}%
\providecommand \href [0]{\begingroup \@sanitize@url \@href}%
\providecommand \@href[1]{\@@startlink{#1}\@@href}%
\providecommand \@@href[1]{\endgroup#1\@@endlink}%
\providecommand \@sanitize@url [0]{\catcode `\\12\catcode `\$12\catcode
  `\&12\catcode `\#12\catcode `\^12\catcode `\_12\catcode `\%12\relax}%
\providecommand \@@startlink[1]{}%
\providecommand \@@endlink[0]{}%
\providecommand \url  [0]{\begingroup\@sanitize@url \@url }%
\providecommand \@url [1]{\endgroup\@href {#1}{\urlprefix }}%
\providecommand \urlprefix  [0]{URL }%
\providecommand \Eprint [0]{\href }%
\providecommand \doibase [0]{http://dx.doi.org/}%
\providecommand \selectlanguage [0]{\@gobble}%
\providecommand \bibinfo  [0]{\@secondoftwo}%
\providecommand \bibfield  [0]{\@secondoftwo}%
\providecommand \translation [1]{[#1]}%
\providecommand \BibitemOpen [0]{}%
\providecommand \bibitemStop [0]{}%
\providecommand \bibitemNoStop [0]{.\EOS\space}%
\providecommand \EOS [0]{\spacefactor3000\relax}%
\providecommand \BibitemShut  [1]{\csname bibitem#1\endcsname}%
\let\auto@bib@innerbib\@empty
\bibitem [{\citenamefont {Schuster}\ \emph {et~al.}(2010)\citenamefont
  {Schuster}, \citenamefont {Sears}, \citenamefont {Ginossar}, \citenamefont
  {DiCarlo}, \citenamefont {Frunzio}, \citenamefont {Morton}, \citenamefont
  {Wu}, \citenamefont {Briggs}, \citenamefont {Buckley}, \citenamefont
  {Awschalom},\ and\ \citenamefont {Schoelkopf}}]{Schuster2010}%
  \BibitemOpen
  \bibfield  {author} {\bibinfo {author} {\bibfnamefont {D.~I.}\ \bibnamefont
  {Schuster}}, \bibinfo {author} {\bibfnamefont {A.~P.}\ \bibnamefont {Sears}},
  \bibinfo {author} {\bibfnamefont {E.}~\bibnamefont {Ginossar}}, \bibinfo
  {author} {\bibfnamefont {L.}~\bibnamefont {DiCarlo}}, \bibinfo {author}
  {\bibfnamefont {L.}~\bibnamefont {Frunzio}}, \bibinfo {author} {\bibfnamefont
  {J.~J.~L.}\ \bibnamefont {Morton}}, \bibinfo {author} {\bibfnamefont
  {H.}~\bibnamefont {Wu}}, \bibinfo {author} {\bibfnamefont {G.~A.~D.}\
  \bibnamefont {Briggs}}, \bibinfo {author} {\bibfnamefont {B.~B.}\
  \bibnamefont {Buckley}}, \bibinfo {author} {\bibfnamefont {D.~D.}\
  \bibnamefont {Awschalom}}, \ and\ \bibinfo {author} {\bibfnamefont {R.~J.}\
  \bibnamefont {Schoelkopf}},\ }\href {\doibase 10.1103/PhysRevLett.105.140501}
  {\bibfield  {journal} {\bibinfo  {journal} {Physical Review Letters}\
  }\textbf {\bibinfo {volume} {105}},\ \bibinfo {pages} {140501} (\bibinfo
  {year} {2010})}\BibitemShut {NoStop}%
\bibitem [{\citenamefont {Kubo}\ \emph {et~al.}(2010)\citenamefont {Kubo},
  \citenamefont {Ong}, \citenamefont {Bertet}, \citenamefont {Vion},
  \citenamefont {Jacques}, \citenamefont {Zheng}, \citenamefont {Dr{\'{e}}au},
  \citenamefont {Roch}, \citenamefont {Auffeves}, \citenamefont {Jelezko},
  \citenamefont {Wrachtrup}, \citenamefont {Barthe}, \citenamefont {Bergonzo},\
  and\ \citenamefont {Esteve}}]{Kubo2010}%
  \BibitemOpen
  \bibfield  {author} {\bibinfo {author} {\bibfnamefont {Y.}~\bibnamefont
  {Kubo}}, \bibinfo {author} {\bibfnamefont {F.~R.}\ \bibnamefont {Ong}},
  \bibinfo {author} {\bibfnamefont {P.}~\bibnamefont {Bertet}}, \bibinfo
  {author} {\bibfnamefont {D.}~\bibnamefont {Vion}}, \bibinfo {author}
  {\bibfnamefont {V.}~\bibnamefont {Jacques}}, \bibinfo {author} {\bibfnamefont
  {D.}~\bibnamefont {Zheng}}, \bibinfo {author} {\bibfnamefont
  {A.}~\bibnamefont {Dr{\'{e}}au}}, \bibinfo {author} {\bibfnamefont {J.-F.}\
  \bibnamefont {Roch}}, \bibinfo {author} {\bibfnamefont {A.}~\bibnamefont
  {Auffeves}}, \bibinfo {author} {\bibfnamefont {F.}~\bibnamefont {Jelezko}},
  \bibinfo {author} {\bibfnamefont {J.}~\bibnamefont {Wrachtrup}}, \bibinfo
  {author} {\bibfnamefont {M.~F.}\ \bibnamefont {Barthe}}, \bibinfo {author}
  {\bibfnamefont {P.}~\bibnamefont {Bergonzo}}, \ and\ \bibinfo {author}
  {\bibfnamefont {D.}~\bibnamefont {Esteve}},\ }\href {\doibase
  10.1103/PhysRevLett.105.140502} {\bibfield  {journal} {\bibinfo  {journal}
  {Physical Review Letters}\ }\textbf {\bibinfo {volume} {105}},\ \bibinfo
  {pages} {140502} (\bibinfo {year} {2010})}\BibitemShut {NoStop}%
\bibitem [{\citenamefont {Zollitsch}\ \emph {et~al.}(2015)\citenamefont
  {Zollitsch}, \citenamefont {Mueller}, \citenamefont {Franke}, \citenamefont
  {Goennenwein}, \citenamefont {Brandt}, \citenamefont {Gross},\ and\
  \citenamefont {Huebl}}]{Zollitsch2015}%
  \BibitemOpen
  \bibfield  {author} {\bibinfo {author} {\bibfnamefont {C.~W.}\ \bibnamefont
  {Zollitsch}}, \bibinfo {author} {\bibfnamefont {K.}~\bibnamefont {Mueller}},
  \bibinfo {author} {\bibfnamefont {D.~P.}\ \bibnamefont {Franke}}, \bibinfo
  {author} {\bibfnamefont {S.~T.~B.}\ \bibnamefont {Goennenwein}}, \bibinfo
  {author} {\bibfnamefont {M.~S.}\ \bibnamefont {Brandt}}, \bibinfo {author}
  {\bibfnamefont {R.}~\bibnamefont {Gross}}, \ and\ \bibinfo {author}
  {\bibfnamefont {H.}~\bibnamefont {Huebl}},\ }\href {\doibase
  10.1063/1.4932658} {\bibfield  {journal} {\bibinfo  {journal} {Applied
  Physics Letters}\ }\textbf {\bibinfo {volume} {107}},\ \bibinfo {pages}
  {142105} (\bibinfo {year} {2015})}\BibitemShut {NoStop}%
\bibitem [{\citenamefont {Soykal}\ and\ \citenamefont
  {Flatt\'e}(2010)}]{Soykal2010}%
  \BibitemOpen
  \bibfield  {author} {\bibinfo {author} {\bibfnamefont {O.~O.}\ \bibnamefont
  {Soykal}}\ and\ \bibinfo {author} {\bibfnamefont {M.~E.}\ \bibnamefont
  {Flatt\'e}},\ }\href {\doibase 10.1103/PhysRevLett.104.077202} {\bibfield
  {journal} {\bibinfo  {journal} {Phys. Rev. Lett.}\ }\textbf {\bibinfo
  {volume} {104}},\ \bibinfo {pages} {077202} (\bibinfo {year}
  {2010})}\BibitemShut {NoStop}%
\bibitem [{\citenamefont {Huebl}\ \emph {et~al.}(2013)\citenamefont {Huebl},
  \citenamefont {Zollitsch}, \citenamefont {Lotze}, \citenamefont {Hocke},
  \citenamefont {Greifenstein}, \citenamefont {Marx}, \citenamefont {Gross},\
  and\ \citenamefont {Goennenwein}}]{Huebl2013}%
  \BibitemOpen
  \bibfield  {author} {\bibinfo {author} {\bibfnamefont {H.}~\bibnamefont
  {Huebl}}, \bibinfo {author} {\bibfnamefont {C.~W.}\ \bibnamefont
  {Zollitsch}}, \bibinfo {author} {\bibfnamefont {J.}~\bibnamefont {Lotze}},
  \bibinfo {author} {\bibfnamefont {F.}~\bibnamefont {Hocke}}, \bibinfo
  {author} {\bibfnamefont {M.}~\bibnamefont {Greifenstein}}, \bibinfo {author}
  {\bibfnamefont {A.}~\bibnamefont {Marx}}, \bibinfo {author} {\bibfnamefont
  {R.}~\bibnamefont {Gross}}, \ and\ \bibinfo {author} {\bibfnamefont
  {S.~T.~B.}\ \bibnamefont {Goennenwein}},\ }\href {\doibase
  10.1103/PhysRevLett.111.127003} {\bibfield  {journal} {\bibinfo  {journal}
  {Physical Review Letters}\ }\textbf {\bibinfo {volume} {111}},\ \bibinfo
  {pages} {127003} (\bibinfo {year} {2013})}\BibitemShut {NoStop}%
\bibitem [{\citenamefont {Tabuchi}\ \emph {et~al.}(2014)\citenamefont
  {Tabuchi}, \citenamefont {Ishino}, \citenamefont {Ishikawa}, \citenamefont
  {Yamazaki}, \citenamefont {Usami},\ and\ \citenamefont
  {Nakamura}}]{Tabuchi2014}%
  \BibitemOpen
  \bibfield  {author} {\bibinfo {author} {\bibfnamefont {Y.}~\bibnamefont
  {Tabuchi}}, \bibinfo {author} {\bibfnamefont {S.}~\bibnamefont {Ishino}},
  \bibinfo {author} {\bibfnamefont {T.}~\bibnamefont {Ishikawa}}, \bibinfo
  {author} {\bibfnamefont {R.}~\bibnamefont {Yamazaki}}, \bibinfo {author}
  {\bibfnamefont {K.}~\bibnamefont {Usami}}, \ and\ \bibinfo {author}
  {\bibfnamefont {Y.}~\bibnamefont {Nakamura}},\ }\href {\doibase
  10.1103/PhysRevLett.113.083603} {\bibfield  {journal} {\bibinfo  {journal}
  {Physical Review Letters}\ }\textbf {\bibinfo {volume} {113}},\ \bibinfo
  {pages} {083603} (\bibinfo {year} {2014})}\BibitemShut {NoStop}%
\bibitem [{\citenamefont {Zhang}\ \emph {et~al.}(2014)\citenamefont {Zhang},
  \citenamefont {Zou}, \citenamefont {Jiang},\ and\ \citenamefont
  {Tang}}]{Zhang2014}%
  \BibitemOpen
  \bibfield  {author} {\bibinfo {author} {\bibfnamefont {X.}~\bibnamefont
  {Zhang}}, \bibinfo {author} {\bibfnamefont {C.-l.}\ \bibnamefont {Zou}},
  \bibinfo {author} {\bibfnamefont {L.}~\bibnamefont {Jiang}}, \ and\ \bibinfo
  {author} {\bibfnamefont {H.~X.}\ \bibnamefont {Tang}},\ }\href {\doibase
  10.1103/PhysRevLett.113.156401} {\bibfield  {journal} {\bibinfo  {journal}
  {Physical Review Letters}\ }\textbf {\bibinfo {volume} {2}},\ \bibinfo
  {pages} {5} (\bibinfo {year} {2014})}\BibitemShut {NoStop}%
\bibitem [{\citenamefont {Zhang}\ \emph {et~al.}(2015)\citenamefont {Zhang},
  \citenamefont {Zou}, \citenamefont {Zhu}, \citenamefont {Marquardt},
  \citenamefont {Jiang},\ and\ \citenamefont {Tang}}]{Zhang2015}%
  \BibitemOpen
  \bibfield  {author} {\bibinfo {author} {\bibfnamefont {X.}~\bibnamefont
  {Zhang}}, \bibinfo {author} {\bibfnamefont {C.-l.}\ \bibnamefont {Zou}},
  \bibinfo {author} {\bibfnamefont {N.}~\bibnamefont {Zhu}}, \bibinfo {author}
  {\bibfnamefont {F.}~\bibnamefont {Marquardt}}, \bibinfo {author}
  {\bibfnamefont {L.}~\bibnamefont {Jiang}}, \ and\ \bibinfo {author}
  {\bibfnamefont {H.~X.}\ \bibnamefont {Tang}},\ }\href {\doibase
  10.1038/ncomms9914} {\bibfield  {journal} {\bibinfo  {journal} {Nature
  Communications}\ }\textbf {\bibinfo {volume} {6}},\ \bibinfo {pages} {8914}
  (\bibinfo {year} {2015})}\BibitemShut {NoStop}%
\bibitem [{\citenamefont {Lambert}\ \emph {et~al.}(2015)\citenamefont
  {Lambert}, \citenamefont {Langenfeld}, \citenamefont {Ferguson},
  \citenamefont {Haigh},\ and\ \citenamefont {Doherty}}]{Lambert2015}%
  \BibitemOpen
  \bibfield  {author} {\bibinfo {author} {\bibfnamefont {N.~J.}\ \bibnamefont
  {Lambert}}, \bibinfo {author} {\bibfnamefont {S.}~\bibnamefont {Langenfeld}},
  \bibinfo {author} {\bibfnamefont {a.~J.}\ \bibnamefont {Ferguson}}, \bibinfo
  {author} {\bibfnamefont {J.~a.}\ \bibnamefont {Haigh}}, \ and\ \bibinfo
  {author} {\bibfnamefont {A.}~\bibnamefont {Doherty}},\ }\href@noop {}
  {\bibfield  {journal} {\bibinfo  {journal} {arXiv preprint}\ } (\bibinfo
  {year} {2015})},\ \Eprint {http://arxiv.org/abs/1506.06049v2}
  {arXiv:1506.06049v2} \BibitemShut {NoStop}%
\bibitem [{\citenamefont {Yao}\ \emph {et~al.}(2015)\citenamefont {Yao},
  \citenamefont {Gui}, \citenamefont {Xiao}, \citenamefont {Guo}, \citenamefont
  {Chen}, \citenamefont {Lu}, \citenamefont {Chien},\ and\ \citenamefont
  {Hu}}]{Yao2015}%
  \BibitemOpen
  \bibfield  {author} {\bibinfo {author} {\bibfnamefont {B.~M.}\ \bibnamefont
  {Yao}}, \bibinfo {author} {\bibfnamefont {Y.~S.}\ \bibnamefont {Gui}},
  \bibinfo {author} {\bibfnamefont {Y.}~\bibnamefont {Xiao}}, \bibinfo {author}
  {\bibfnamefont {H.}~\bibnamefont {Guo}}, \bibinfo {author} {\bibfnamefont
  {X.~S.}\ \bibnamefont {Chen}}, \bibinfo {author} {\bibfnamefont
  {W.}~\bibnamefont {Lu}}, \bibinfo {author} {\bibfnamefont {C.~L.}\
  \bibnamefont {Chien}}, \ and\ \bibinfo {author} {\bibfnamefont {C.-M.}\
  \bibnamefont {Hu}},\ }\href {\doibase 10.1103/PhysRevB.92.184407} {\bibfield
  {journal} {\bibinfo  {journal} {Physical Review B}\ }\textbf {\bibinfo
  {volume} {92}},\ \bibinfo {pages} {184407} (\bibinfo {year}
  {2015})}\BibitemShut {NoStop}%
\bibitem [{\citenamefont {Czeschka}\ \emph {et~al.}(2011)\citenamefont
  {Czeschka}, \citenamefont {Dreher}, \citenamefont {Brandt}, \citenamefont
  {Weiler}, \citenamefont {Althammer}, \citenamefont {Imort}, \citenamefont
  {Reiss}, \citenamefont {Thomas}, \citenamefont {Schoch}, \citenamefont
  {Limmer}, \citenamefont {Huebl}, \citenamefont {Gross},\ and\ \citenamefont
  {Goennenwein}}]{Czeschka2011a}%
  \BibitemOpen
  \bibfield  {author} {\bibinfo {author} {\bibfnamefont {F.~D.}\ \bibnamefont
  {Czeschka}}, \bibinfo {author} {\bibfnamefont {L.}~\bibnamefont {Dreher}},
  \bibinfo {author} {\bibfnamefont {M.~S.}\ \bibnamefont {Brandt}}, \bibinfo
  {author} {\bibfnamefont {M.}~\bibnamefont {Weiler}}, \bibinfo {author}
  {\bibfnamefont {M.}~\bibnamefont {Althammer}}, \bibinfo {author}
  {\bibfnamefont {I.-M.}\ \bibnamefont {Imort}}, \bibinfo {author}
  {\bibfnamefont {G.}~\bibnamefont {Reiss}}, \bibinfo {author} {\bibfnamefont
  {A.}~\bibnamefont {Thomas}}, \bibinfo {author} {\bibfnamefont
  {W.}~\bibnamefont {Schoch}}, \bibinfo {author} {\bibfnamefont
  {W.}~\bibnamefont {Limmer}}, \bibinfo {author} {\bibfnamefont
  {H.}~\bibnamefont {Huebl}}, \bibinfo {author} {\bibfnamefont
  {R.}~\bibnamefont {Gross}}, \ and\ \bibinfo {author} {\bibfnamefont
  {S.~T.~B.}\ \bibnamefont {Goennenwein}},\ }\href {\doibase
  10.1103/PhysRevLett.107.046601} {\bibfield  {journal} {\bibinfo  {journal}
  {Physical Review Letters}\ }\textbf {\bibinfo {volume} {107}},\ \bibinfo
  {pages} {046601} (\bibinfo {year} {2011})}\BibitemShut {NoStop}%
\bibitem [{\citenamefont {Iguchi}\ \emph {et~al.}(2014)\citenamefont {Iguchi},
  \citenamefont {Sato}, \citenamefont {Hirobe}, \citenamefont {Daimon},\ and\
  \citenamefont {Saitoh}}]{Iguchi2014}%
  \BibitemOpen
  \bibfield  {author} {\bibinfo {author} {\bibfnamefont {R.}~\bibnamefont
  {Iguchi}}, \bibinfo {author} {\bibfnamefont {K.}~\bibnamefont {Sato}},
  \bibinfo {author} {\bibfnamefont {D.}~\bibnamefont {Hirobe}}, \bibinfo
  {author} {\bibfnamefont {S.}~\bibnamefont {Daimon}}, \ and\ \bibinfo {author}
  {\bibfnamefont {E.}~\bibnamefont {Saitoh}},\ }\href {\doibase
  10.7567/APEX.7.013003} {\bibfield  {journal} {\bibinfo  {journal} {Applied
  Physics Express}\ }\textbf {\bibinfo {volume} {7}},\ \bibinfo {pages}
  {013003} (\bibinfo {year} {2014})}\BibitemShut {NoStop}%
\bibitem [{\citenamefont {Tserkovnyak}\ \emph {et~al.}(2002)\citenamefont
  {Tserkovnyak}, \citenamefont {Brataas},\ and\ \citenamefont
  {Bauer}}]{Tserkovnyak2002}%
  \BibitemOpen
  \bibfield  {author} {\bibinfo {author} {\bibfnamefont {Y.}~\bibnamefont
  {Tserkovnyak}}, \bibinfo {author} {\bibfnamefont {A.}~\bibnamefont
  {Brataas}}, \ and\ \bibinfo {author} {\bibfnamefont {G.~E.~W.}\ \bibnamefont
  {Bauer}},\ }\href {\doibase 10.1103/PhysRevLett.88.117601} {\bibfield
  {journal} {\bibinfo  {journal} {Physical Review Letters}\ }\textbf {\bibinfo
  {volume} {88}},\ \bibinfo {pages} {117601} (\bibinfo {year}
  {2002})}\BibitemShut {NoStop}%
\bibitem [{\citenamefont {Bai}\ \emph {et~al.}(2015)\citenamefont {Bai},
  \citenamefont {Harder}, \citenamefont {Chen}, \citenamefont {Fan},
  \citenamefont {Xiao},\ and\ \citenamefont {Hu}}]{Bai2015}%
  \BibitemOpen
  \bibfield  {author} {\bibinfo {author} {\bibfnamefont {L.}~\bibnamefont
  {Bai}}, \bibinfo {author} {\bibfnamefont {M.}~\bibnamefont {Harder}},
  \bibinfo {author} {\bibfnamefont {Y.~P.}\ \bibnamefont {Chen}}, \bibinfo
  {author} {\bibfnamefont {X.}~\bibnamefont {Fan}}, \bibinfo {author}
  {\bibfnamefont {J.~Q.}\ \bibnamefont {Xiao}}, \ and\ \bibinfo {author}
  {\bibfnamefont {C.-M.}\ \bibnamefont {Hu}},\ }\href {\doibase
  10.1103/PhysRevLett.114.227201} {\bibfield  {journal} {\bibinfo  {journal}
  {Physical Review Letters}\ }\textbf {\bibinfo {volume} {114}},\ \bibinfo
  {pages} {227201} (\bibinfo {year} {2015})}\BibitemShut {NoStop}%
\bibitem [{\citenamefont {Cao}\ \emph {et~al.}(2014)\citenamefont {Cao},
  \citenamefont {Yan}, \citenamefont {Huebl}, \citenamefont {Goennenwein},\
  and\ \citenamefont {Bauer}}]{Cao2015}%
  \BibitemOpen
  \bibfield  {author} {\bibinfo {author} {\bibfnamefont {Y.}~\bibnamefont
  {Cao}}, \bibinfo {author} {\bibfnamefont {P.}~\bibnamefont {Yan}}, \bibinfo
  {author} {\bibfnamefont {H.}~\bibnamefont {Huebl}}, \bibinfo {author}
  {\bibfnamefont {S.~T.~B.}\ \bibnamefont {Goennenwein}}, \ and\ \bibinfo
  {author} {\bibfnamefont {G.~E.~W.}\ \bibnamefont {Bauer}},\ }\href {\doibase
  10.1103/PhysRevB.91.094423} {\bibfield  {journal} {\bibinfo  {journal}
  {Physical Review B}\ }\textbf {\bibinfo {volume} {094423}},\ \bibinfo {pages}
  {5} (\bibinfo {year} {2014})}\BibitemShut {NoStop}%
\bibitem [{\citenamefont {Lotze}(2015)}]{Lotze2015}%
  \BibitemOpen
  \bibfield  {author} {\bibinfo {author} {\bibfnamefont {J.}~\bibnamefont
  {Lotze}},\ }\emph {\bibinfo {title} {Spin Pumping in Ferrimagnet / Normal
  Metal Bilayers}},\ \href@noop {} {Ph.D. thesis},\ \bibinfo  {school}
  {Technische Universit{\"{a}}t M{\"{u}}nchen} (\bibinfo {year}
  {2015})\BibitemShut {NoStop}%
\bibitem [{\citenamefont {Brandlmaier}\ \emph {et~al.}(2008)\citenamefont
  {Brandlmaier}, \citenamefont {Gepr{\"{a}}gs}, \citenamefont {Weiler},
  \citenamefont {Boger}, \citenamefont {Opel}, \citenamefont {Huebl},
  \citenamefont {Bihler}, \citenamefont {Brandt}, \citenamefont {Botters},
  \citenamefont {Grundler}, \citenamefont {Gross},\ and\ \citenamefont
  {Goennenwein}}]{Brandlmaier2008}%
  \BibitemOpen
  \bibfield  {author} {\bibinfo {author} {\bibfnamefont {A.}~\bibnamefont
  {Brandlmaier}}, \bibinfo {author} {\bibfnamefont {S.}~\bibnamefont
  {Gepr{\"{a}}gs}}, \bibinfo {author} {\bibfnamefont {M.}~\bibnamefont
  {Weiler}}, \bibinfo {author} {\bibfnamefont {A.}~\bibnamefont {Boger}},
  \bibinfo {author} {\bibfnamefont {M.}~\bibnamefont {Opel}}, \bibinfo {author}
  {\bibfnamefont {H.}~\bibnamefont {Huebl}}, \bibinfo {author} {\bibfnamefont
  {C.}~\bibnamefont {Bihler}}, \bibinfo {author} {\bibfnamefont {M.~S.}\
  \bibnamefont {Brandt}}, \bibinfo {author} {\bibfnamefont {B.}~\bibnamefont
  {Botters}}, \bibinfo {author} {\bibfnamefont {D.}~\bibnamefont {Grundler}},
  \bibinfo {author} {\bibfnamefont {R.}~\bibnamefont {Gross}}, \ and\ \bibinfo
  {author} {\bibfnamefont {S.~T.~B.}\ \bibnamefont {Goennenwein}},\ }\href
  {\doibase 10.1103/PhysRevB.77.104445} {\bibfield  {journal} {\bibinfo
  {journal} {Physical Review B}\ }\textbf {\bibinfo {volume} {77}},\ \bibinfo
  {pages} {1} (\bibinfo {year} {2008})}\BibitemShut {NoStop}%
\bibitem [{\citenamefont {Heinrich}\ \emph {et~al.}(2003)\citenamefont
  {Heinrich}, \citenamefont {Tserkovnyak}, \citenamefont {Woltersdorf},
  \citenamefont {Brataas}, \citenamefont {Urban},\ and\ \citenamefont
  {Bauer}}]{Heinrich2003}%
  \BibitemOpen
  \bibfield  {author} {\bibinfo {author} {\bibfnamefont {B.}~\bibnamefont
  {Heinrich}}, \bibinfo {author} {\bibfnamefont {Y.}~\bibnamefont
  {Tserkovnyak}}, \bibinfo {author} {\bibfnamefont {G.}~\bibnamefont
  {Woltersdorf}}, \bibinfo {author} {\bibfnamefont {A.}~\bibnamefont
  {Brataas}}, \bibinfo {author} {\bibfnamefont {R.}~\bibnamefont {Urban}}, \
  and\ \bibinfo {author} {\bibfnamefont {G.~E.~W.}\ \bibnamefont {Bauer}},\
  }\href {\doibase 10.1103/PhysRevLett.90.187601} {\bibfield  {journal}
  {\bibinfo  {journal} {Physical Review Letters}\ }\textbf {\bibinfo {volume}
  {90}},\ \bibinfo {pages} {187601} (\bibinfo {year} {2003})}\BibitemShut
  {NoStop}%
\bibitem [{\citenamefont {Mosendz}\ \emph {et~al.}(2010)\citenamefont
  {Mosendz}, \citenamefont {Pearson}, \citenamefont {Fradin}, \citenamefont
  {Bauer}, \citenamefont {Bader},\ and\ \citenamefont
  {Hoffmann}}]{Mosendz2010}%
  \BibitemOpen
  \bibfield  {author} {\bibinfo {author} {\bibfnamefont {O.}~\bibnamefont
  {Mosendz}}, \bibinfo {author} {\bibfnamefont {J.~E.}\ \bibnamefont
  {Pearson}}, \bibinfo {author} {\bibfnamefont {F.~Y.}\ \bibnamefont {Fradin}},
  \bibinfo {author} {\bibfnamefont {G.~E.~W.}\ \bibnamefont {Bauer}}, \bibinfo
  {author} {\bibfnamefont {S.~D.}\ \bibnamefont {Bader}}, \ and\ \bibinfo
  {author} {\bibfnamefont {A.}~\bibnamefont {Hoffmann}},\ }\href {\doibase
  10.1103/PhysRevLett.104.046601} {\bibfield  {journal} {\bibinfo  {journal}
  {Physical Review Letters}\ }\textbf {\bibinfo {volume} {104}},\ \bibinfo
  {pages} {046601} (\bibinfo {year} {2010})}\BibitemShut {NoStop}%
\bibitem [{\citenamefont {Kittel}(1995)}]{Kittel2005}%
  \BibitemOpen
  \bibfield  {author} {\bibinfo {author} {\bibfnamefont {C.}~\bibnamefont
  {Kittel}},\ }\href@noop {} {\emph {\bibinfo {title} {Introduction to Solid
  State Physics}}}\ (\bibinfo  {publisher} {John Wiley {\&} Sons},\ \bibinfo
  {address} {New York},\ \bibinfo {year} {1995})\BibitemShut {NoStop}%
\bibitem [{\citenamefont {Tavis}\ and\ \citenamefont
  {Cummings}(1968)}]{Tavis1968}%
  \BibitemOpen
  \bibfield  {author} {\bibinfo {author} {\bibfnamefont {M.}~\bibnamefont
  {Tavis}}\ and\ \bibinfo {author} {\bibfnamefont {F.~W.}\ \bibnamefont
  {Cummings}},\ }\href {\doibase 10.1103/PhysRev.170.379} {\bibfield  {journal}
  {\bibinfo  {journal} {Physical Review}\ }\textbf {\bibinfo {volume} {170}},\
  \bibinfo {pages} {379} (\bibinfo {year} {1968})}\BibitemShut {NoStop}%
\bibitem [{\citenamefont {Fink}\ \emph {et~al.}(2009)\citenamefont {Fink},
  \citenamefont {Bianchetti}, \citenamefont {Baur}, \citenamefont
  {G{\"{o}}ppl}, \citenamefont {Steffen}, \citenamefont {Filipp}, \citenamefont
  {Leek}, \citenamefont {Blais},\ and\ \citenamefont {Wallraff}}]{Fink2009}%
  \BibitemOpen
  \bibfield  {author} {\bibinfo {author} {\bibfnamefont {J.~M.}\ \bibnamefont
  {Fink}}, \bibinfo {author} {\bibfnamefont {R.}~\bibnamefont {Bianchetti}},
  \bibinfo {author} {\bibfnamefont {M.}~\bibnamefont {Baur}}, \bibinfo {author}
  {\bibfnamefont {M.}~\bibnamefont {G{\"{o}}ppl}}, \bibinfo {author}
  {\bibfnamefont {L.}~\bibnamefont {Steffen}}, \bibinfo {author} {\bibfnamefont
  {S.}~\bibnamefont {Filipp}}, \bibinfo {author} {\bibfnamefont {P.~J.}\
  \bibnamefont {Leek}}, \bibinfo {author} {\bibfnamefont {A.}~\bibnamefont
  {Blais}}, \ and\ \bibinfo {author} {\bibfnamefont {A.}~\bibnamefont
  {Wallraff}},\ }\href {\doibase 10.1103/PhysRevLett.103.083601} {\bibfield
  {journal} {\bibinfo  {journal} {Physical Review Letters}\ }\textbf {\bibinfo
  {volume} {103}},\ \bibinfo {pages} {083601} (\bibinfo {year}
  {2009})}\BibitemShut {NoStop}%
\bibitem [{\citenamefont {Jungfleisch}\ \emph {et~al.}(2013)\citenamefont
  {Jungfleisch}, \citenamefont {Lauer}, \citenamefont {Neb}, \citenamefont
  {Chumak},\ and\ \citenamefont {Hillebrands}}]{Jungfleisch2013}%
  \BibitemOpen
  \bibfield  {author} {\bibinfo {author} {\bibfnamefont {M.~B.}\ \bibnamefont
  {Jungfleisch}}, \bibinfo {author} {\bibfnamefont {V.}~\bibnamefont {Lauer}},
  \bibinfo {author} {\bibfnamefont {R.}~\bibnamefont {Neb}}, \bibinfo {author}
  {\bibfnamefont {a.~V.}\ \bibnamefont {Chumak}}, \ and\ \bibinfo {author}
  {\bibfnamefont {B.}~\bibnamefont {Hillebrands}},\ }\href {\doibase
  10.1063/1.4813315} {\bibfield  {journal} {\bibinfo  {journal} {Applied
  Physics Letters}\ }\textbf {\bibinfo {volume} {103}},\ \bibinfo {pages}
  {2011} (\bibinfo {year} {2013})}\BibitemShut {NoStop}%
\bibitem [{\citenamefont {Gilleo}\ and\ \citenamefont
  {Geller}(1958)}]{Gilleo1958}%
  \BibitemOpen
  \bibfield  {author} {\bibinfo {author} {\bibfnamefont {M.~A.}\ \bibnamefont
  {Gilleo}}\ and\ \bibinfo {author} {\bibfnamefont {S.}~\bibnamefont
  {Geller}},\ }\href {\doibase 10.1103/PhysRev.110.73} {\bibfield  {journal}
  {\bibinfo  {journal} {Physical Review}\ }\textbf {\bibinfo {volume} {110}},\
  \bibinfo {pages} {73} (\bibinfo {year} {1958})}\BibitemShut {NoStop}%
\bibitem [{Note1()}]{Note1}%
  \BibitemOpen
  \bibinfo {note} {The designs of the sample holder are published under the
  CERN Open Hardware License \protect \texttt
  {http://hannes.maier-flaig.de/flexline-sample-rod}}\BibitemShut {NoStop}%
\bibitem [{\citenamefont {Chiorescu}\ \emph {et~al.}(2010)\citenamefont
  {Chiorescu}, \citenamefont {Groll}, \citenamefont {Bertaina}, \citenamefont
  {Mori},\ and\ \citenamefont {Miyashita}}]{Chiorescu2010}%
  \BibitemOpen
  \bibfield  {author} {\bibinfo {author} {\bibfnamefont {I.}~\bibnamefont
  {Chiorescu}}, \bibinfo {author} {\bibfnamefont {N.}~\bibnamefont {Groll}},
  \bibinfo {author} {\bibfnamefont {S.}~\bibnamefont {Bertaina}}, \bibinfo
  {author} {\bibfnamefont {T.}~\bibnamefont {Mori}}, \ and\ \bibinfo {author}
  {\bibfnamefont {S.}~\bibnamefont {Miyashita}},\ }\href {\doibase
  10.1103/PhysRevB.82.024413} {\bibfield  {journal} {\bibinfo  {journal}
  {Physical Review B}\ }\textbf {\bibinfo {volume} {82}},\ \bibinfo {pages}
  {024413} (\bibinfo {year} {2010})}\BibitemShut {NoStop}%
\bibitem [{\citenamefont {Petersan}\ and\ \citenamefont
  {Anlage}(1998)}]{Petersan1998}%
  \BibitemOpen
  \bibfield  {author} {\bibinfo {author} {\bibfnamefont {P.~J.}\ \bibnamefont
  {Petersan}}\ and\ \bibinfo {author} {\bibfnamefont {S.~M.}\ \bibnamefont
  {Anlage}},\ }\href {\doibase 10.1063/1.368498} {\bibfield  {journal}
  {\bibinfo  {journal} {Journal of Applied Physics}\ }\textbf {\bibinfo
  {volume} {84}},\ \bibinfo {pages} {3392} (\bibinfo {year}
  {1998})}\BibitemShut {NoStop}%
\bibitem [{\citenamefont {Abe}\ \emph {et~al.}(2011)\citenamefont {Abe},
  \citenamefont {Wu}, \citenamefont {Ardavan},\ and\ \citenamefont
  {Morton}}]{Abe2011}%
  \BibitemOpen
  \bibfield  {author} {\bibinfo {author} {\bibfnamefont {E.}~\bibnamefont
  {Abe}}, \bibinfo {author} {\bibfnamefont {H.}~\bibnamefont {Wu}}, \bibinfo
  {author} {\bibfnamefont {A.}~\bibnamefont {Ardavan}}, \ and\ \bibinfo
  {author} {\bibfnamefont {J.~J.~L.}\ \bibnamefont {Morton}},\ }\href {\doibase
  10.1063/1.3601930} {\bibfield  {journal} {\bibinfo  {journal} {Applied
  Physics Letters}\ }\textbf {\bibinfo {volume} {98}},\ \bibinfo {pages}
  {251108} (\bibinfo {year} {2011})}\BibitemShut {NoStop}%
\bibitem [{\citenamefont {Goennenwein}\ \emph {et~al.}(2003)\citenamefont
  {Goennenwein}, \citenamefont {Graf}, \citenamefont {Wassner}, \citenamefont
  {Brandt}, \citenamefont {Stutzmann}, \citenamefont {Philipp}, \citenamefont
  {Gross}, \citenamefont {Ziemann}, \citenamefont {Krieger}, \citenamefont
  {Zu}, \citenamefont {Koeder}, \citenamefont {Frank}, \citenamefont {Schoch},\
  and\ \citenamefont {Waag}}]{Goennenwein2003}%
  \BibitemOpen
  \bibfield  {author} {\bibinfo {author} {\bibfnamefont {S.~T.~B.}\
  \bibnamefont {Goennenwein}}, \bibinfo {author} {\bibfnamefont
  {T.}~\bibnamefont {Graf}}, \bibinfo {author} {\bibfnamefont {T.}~\bibnamefont
  {Wassner}}, \bibinfo {author} {\bibfnamefont {M.~S.}\ \bibnamefont {Brandt}},
  \bibinfo {author} {\bibfnamefont {M.}~\bibnamefont {Stutzmann}}, \bibinfo
  {author} {\bibfnamefont {J.~B.}\ \bibnamefont {Philipp}}, \bibinfo {author}
  {\bibfnamefont {R.}~\bibnamefont {Gross}}, \bibinfo {author} {\bibfnamefont
  {P.}~\bibnamefont {Ziemann}}, \bibinfo {author} {\bibfnamefont
  {M.}~\bibnamefont {Krieger}}, \bibinfo {author} {\bibfnamefont
  {K.}~\bibnamefont {Zu}}, \bibinfo {author} {\bibfnamefont {A.}~\bibnamefont
  {Koeder}}, \bibinfo {author} {\bibfnamefont {S.}~\bibnamefont {Frank}},
  \bibinfo {author} {\bibfnamefont {W.}~\bibnamefont {Schoch}}, \ and\ \bibinfo
  {author} {\bibfnamefont {A.}~\bibnamefont {Waag}},\ }\href {\doibase
  10.1063/1.1539550} {\bibfield  {journal} {\bibinfo  {journal} {Applied
  Physics Letters}\ }\textbf {\bibinfo {volume} {82}},\ \bibinfo {pages} {2003}
  (\bibinfo {year} {2003})}\BibitemShut {NoStop}%
\bibitem [{\citenamefont {Herskind}\ \emph {et~al.}(2009)\citenamefont
  {Herskind}, \citenamefont {Dantan}, \citenamefont {Marler}, \citenamefont
  {Albert},\ and\ \citenamefont {Drewsen}}]{Herskind2009}%
  \BibitemOpen
  \bibfield  {author} {\bibinfo {author} {\bibfnamefont {P.~F.}\ \bibnamefont
  {Herskind}}, \bibinfo {author} {\bibfnamefont {A.}~\bibnamefont {Dantan}},
  \bibinfo {author} {\bibfnamefont {J.~P.}\ \bibnamefont {Marler}}, \bibinfo
  {author} {\bibfnamefont {M.}~\bibnamefont {Albert}}, \ and\ \bibinfo {author}
  {\bibfnamefont {M.}~\bibnamefont {Drewsen}},\ }\href {\doibase
  10.1038/nphys1302} {\bibfield  {journal} {\bibinfo  {journal} {Nature
  Physics}\ }\textbf {\bibinfo {volume} {5}},\ \bibinfo {pages} {494} (\bibinfo
  {year} {2009})}\BibitemShut {NoStop}%
\bibitem [{\citenamefont {Hoekstra}\ \emph {et~al.}(1977)\citenamefont
  {Hoekstra}, \citenamefont {van Stapele},\ and\ \citenamefont
  {Robertson}}]{Hoekstra1977}%
  \BibitemOpen
  \bibfield  {author} {\bibinfo {author} {\bibfnamefont {B.}~\bibnamefont
  {Hoekstra}}, \bibinfo {author} {\bibfnamefont {R.~P.}\ \bibnamefont {van
  Stapele}}, \ and\ \bibinfo {author} {\bibfnamefont {J.~M.}\ \bibnamefont
  {Robertson}},\ }\href {\doibase 10.1063/1.323339} {\bibfield  {journal}
  {\bibinfo  {journal} {Journal of Applied Physics}\ }\textbf {\bibinfo
  {volume} {48}},\ \bibinfo {pages} {382} (\bibinfo {year} {1977})}\BibitemShut
  {NoStop}%
\bibitem [{\citenamefont {Tucciarone}\ and\ \citenamefont {{De
  Gasperis}}(1984)}]{Tucciarone1984}%
  \BibitemOpen
  \bibfield  {author} {\bibinfo {author} {\bibfnamefont {A.}~\bibnamefont
  {Tucciarone}}\ and\ \bibinfo {author} {\bibfnamefont {P.}~\bibnamefont {{De
  Gasperis}}},\ }\href {\doibase 10.1016/0040-6090(84)90338-9} {\bibfield
  {journal} {\bibinfo  {journal} {Thin Solid Films}\ }\textbf {\bibinfo
  {volume} {114}},\ \bibinfo {pages} {109} (\bibinfo {year}
  {1984})}\BibitemShut {NoStop}%
\bibitem [{\citenamefont {Kajiwara}\ \emph {et~al.}(2010)\citenamefont
  {Kajiwara}, \citenamefont {Harii}, \citenamefont {Takahashi}, \citenamefont
  {Ohe}, \citenamefont {Uchida}, \citenamefont {Mizuguchi}, \citenamefont
  {Umezawa}, \citenamefont {Kawai}, \citenamefont {Ando}, \citenamefont
  {Takanashi}, \citenamefont {Maekawa},\ and\ \citenamefont
  {Saitoh}}]{Kajiwara2010}%
  \BibitemOpen
  \bibfield  {author} {\bibinfo {author} {\bibfnamefont {Y.}~\bibnamefont
  {Kajiwara}}, \bibinfo {author} {\bibfnamefont {K.}~\bibnamefont {Harii}},
  \bibinfo {author} {\bibfnamefont {S.}~\bibnamefont {Takahashi}}, \bibinfo
  {author} {\bibfnamefont {J.}~\bibnamefont {Ohe}}, \bibinfo {author}
  {\bibfnamefont {K.}~\bibnamefont {Uchida}}, \bibinfo {author} {\bibfnamefont
  {M.}~\bibnamefont {Mizuguchi}}, \bibinfo {author} {\bibfnamefont
  {H.}~\bibnamefont {Umezawa}}, \bibinfo {author} {\bibfnamefont
  {H.}~\bibnamefont {Kawai}}, \bibinfo {author} {\bibfnamefont
  {K.}~\bibnamefont {Ando}}, \bibinfo {author} {\bibfnamefont {K.}~\bibnamefont
  {Takanashi}}, \bibinfo {author} {\bibfnamefont {S.}~\bibnamefont {Maekawa}},
  \ and\ \bibinfo {author} {\bibfnamefont {E.}~\bibnamefont {Saitoh}},\ }\href
  {\doibase 10.1038/nature08876} {\bibfield  {journal} {\bibinfo  {journal}
  {Nature}\ }\textbf {\bibinfo {volume} {464}},\ \bibinfo {pages} {262}
  (\bibinfo {year} {2010})}\BibitemShut {NoStop}%
\end{thebibliography}%

\end{document}